# CCTCOVID: COVID-19 Detection from Chest X-Ray Images Using Compact Convolutional Transformers

Abdolreza Marefat[1], Mahdieh Marefat[2], Javad Hasannataj Joloudari[3], Mohammad Ali Nematollahi[4,*], Reza Lashgari[5]

[1] Department of Computer Engineering, South Tehran Branch, Islamic Azad University, Tehran, Iran
[2] Department of Cellular and Molecular Biology, Science and Research Branch, Islamic Azad University, Tehran, Iran
[3] Department of Computer Engineering, Faculty of Engineering, University of Birjand, Birjand 9717434765, Iran
[4] Department of Computer Sciences, Fasa University, Fasa, Iran
[5] Institute of Medical Science and Technology, Shahid Beheshti University, Tehran, Iran

Corresponding author*: ma.nematollahi@fasau.ac.ir

## Abstract

COVID-19 is a novel virus that attacks the upper respiratory tract and the lungs. Its person-to-person transmissibility is considerably rapid and this has caused serious problems in approximately every facet of individuals' lives. While some infected individuals may remain completely asymptomatic, others have been frequently witnessed to have mild to severe symptoms. In addition to this, thousands of death cases around the globe indicated that detecting COVID-19 is an urgent demand in the communities. Practically, this is prominently done with the help of screening medical images such as Computed Tomography (CT) and X-ray images. However, the cumbersome clinical procedures and a large number of daily cases have imposed great challenges on medical practitioners. Deep Learning-based approaches have demonstrated a profound potential in a wide range of medical tasks. As a result, we introduce a transformer-based method for automatically detecting COVID-19 from X-ray images using Compact Convolutional Transformers (CCT). Our extensive experiments prove the efficacy of the proposed method with an accuracy of 98% which outperforms the previous works.

## 1. Introduction

The virus, named Severe Acute Respiratory Syndrome Corona-Virus 2 SARS-CoV-2, also known by the name of COVID-19, is the source of a severe disease that started in Wuhan, China during the last months of 2019 [1]. It soon spread to other parts of the globe and caused one of the most devastating pandemics, in that millions of people became abruptly affected or dead. According to the World Health Organization (WHO), the number of death cases in the first half of 2022 stood at more than 6,200,000 and the number of diagnosed people reached more than 516,000,000 in the same year worldwide. This virus belongs to the same group as Severe Acute Respiratory Syndrome (SARS) and Middle East Respiratory Syndrome (MERS) [2]. Its commonly recognized symptoms are coughing, shortness of breath, fever, pneumonia, and respiratory distress [3].

The negative ramifications, imposed on the communities by this virus, and also its rapid transmission from one person to another, prove the necessity of tackling this disease with prohibitive measures. Approximately all countries included a variety of safety protocols, such as social distancing, with the object of controlling the outbreak of this pandemic. Accurately and rapidly detecting COVID-19 is an essential step that should be taken to control the widespread disease [4]. Screening and monitoring of Computed Tomography (CT) and X-ray images have demonstrated great potential in providing a reliable modality for experts to examine different lung diseases such as tuberculosis, infiltra-tion, atelectasis, pneumonia, and COVID-19 [5]. However, the lack of specialized human resources in many regions, especially poor and underdeveloped countries acts as an impediment to taking advantage of such imaging technologies. This motivated the scientific community to utilize computer-aided intelligent decision-making systems to automate the required process.

Deep Learning (DL) is a powerful tool that can provide us with such systems. Among various architectures, designed for processing different types of data, Convolutional Neural Networks (CNNs) and Vision Transformers (ViTs) are specifically invented for visual data. Especially, in medical image analysis, these architectures have proven to be remarkably effective for diagnosing a wide variety of medical conditions. In the following, a brief explanation of CNNs and ViTs is given.

## 1.1. Convolutional Neural Network

Convolutional Neural Networks (CNNs) is one of the most favored types of architectures in deep learning, especially in computer vision [6]. The main component of CNN-based architectures is convolution, which is a mathematical linear operation between matrices [7]. CNNs' most notable success is in the field of pattern recognition applied to imagery, that is, visual data [8]. In fact, the introduction of CNNs by [9], has revolutionized a wide variety of challenges in the domain of computer vision such as medical image analysis, face recognition, image classification, object detection, and semantic segmentation [10; 11; 12; 13; 14; 15].

In general, CNN-based models comprise three types of layers, namely convolutional layers, pooling layers, and fully-connected layers. These three are depicted in Figure 1, where you can see a formation of a CNN-based model for classifying the input lung X-ray image into healthy or unhealthy samples. As is shown in this figure, the convolution layer operates by sliding a kernel on the input data. Each kernel results in a feature map, to which the pooling operation is applied.

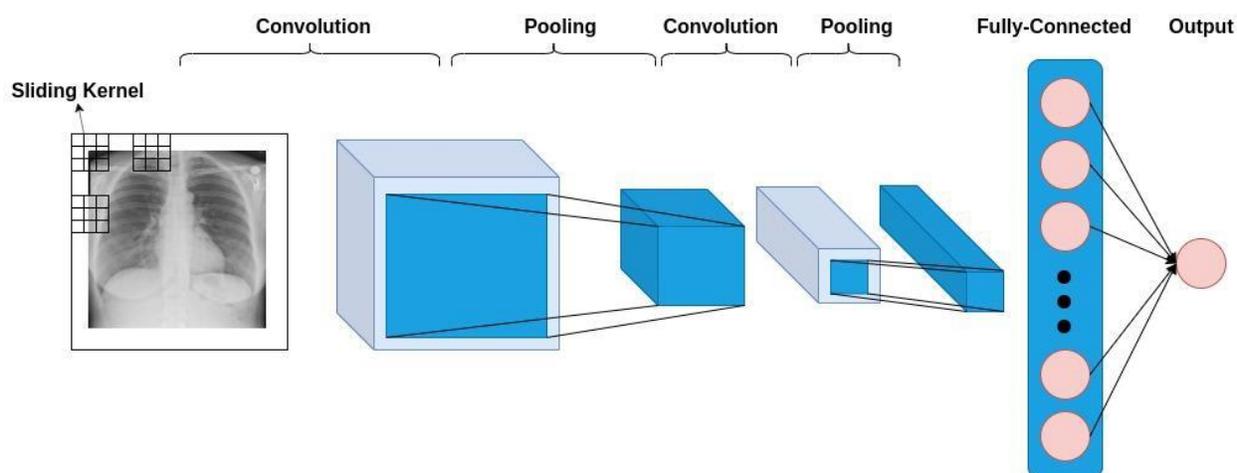

**Figure 1.** The general architecture of a CNN-based model.

Furthermore, translation equivariance and translational invariance, which is inherent to CNNs, enable them to learn natural statistics of the input image. In addition to this, sparse interaction, weight sharing, and equivariant representations make CNN-based models more efficient and less computationally expensive [16].

## 1.2. Vision Transformer

Transformer-based models in deep neural networks have been originally introduced in the domain of Natural Language Processing (NLP) [17]. The astounding performance of these models in a variety of tasks in NLP, i.e., machine translation [18], question answering [19], text classification [20], sentiment analysis [20; 21], has sparked the interest of a considerable number of researchers in computer vision to attune these models to the field of computer vision [22]. [23] was the first research paper, in which the authors creatively invented a way to apply transformers to the visual data and introduced ViTs for image classification. Figure 2 demonstrates a general procedure in ViT-based models. Based on this figure, it can be witnessed that an image is converted to a set of patches, each representing a locality of a region in the image. This procedure enables us to look upon an image as sequential data; the type of data that is prevalent in NLP and is tailored for transformers.

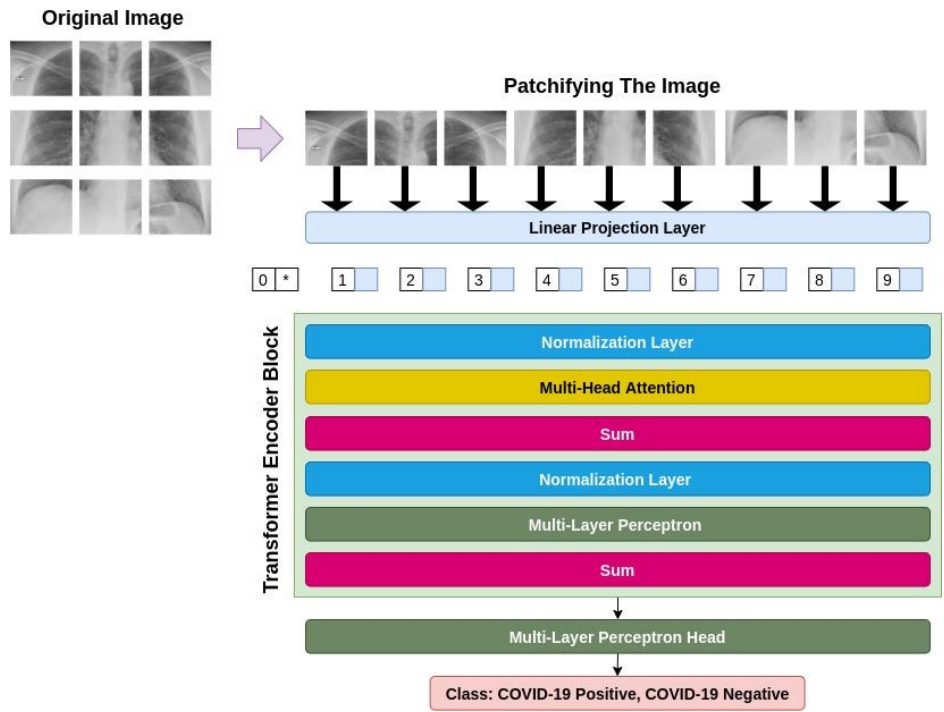

**Figure 2.** The general architecture of a ViT-based model.

Firstly, ViT flattens these patches and then passes them through a trainable linear projection layer, making the projections the same with regard to their dimensionality. Then, since the ViT is thoroughly agnostic to the hierarchy of the input image, meaning that it does not take into consideration where each patch is located in the original image, the position embeddings are integrated into these projections to eradicate this problem. After that, the transformer encoder block takes these patches, alongside their position, and an extra classification token named CLS token. The transformer encoder includes multi-head attention layers, capable of learning a variety of self-attention states. Lastly, the outputs of all existent heads are amalgamated and fed to the Multi-Layer Perceptron (MLP).

## 2. Related works

In this section, we present a brief review of the previous works for detecting COVID-19 from CT or X-Ray images. Due to the successful performance of deep learning-based approaches in medical image analysis [24], researchers have focused on proposing different CNN or ViT-based architectures in order to automatically recognize the presence of the infection [5].

To begin with, Wang et al. [25] were one of the first groups who designed a deep neural network for detecting COVID-19. In addition to this, they provided a relatively large dataset of chest X-ray images. They achieved 93.3 % accuracy. In [26], Marques et al. proposed a pipeline based on EfficientNet and followed the 10-fold cross-validation paradigm to evaluate their approach to chest x-ray images. They have achieved an average accuracy of 99.62% and 97.54% in binary and multi-class classification, respectively. Singh et al [27]. utilized a famous neural network, named VGG16, and transfer learning in order to detect COVID-19 from CT scans. In their approach, the extracted features were chosen by using Principal Component Analysis (PCA) and later classified by different classifiers. At most, they achieved 95.7% accuracy. In [28], Zabirul Islam et al. made a neural network that was a hybrid of CNNs and Long Short-Term Memory (LSTM) networks. They trained their model on 3 classes, namely COVID-19, pneumonia, and normal, and achieved 99.2%, 99.2%, and 99.8% accuracy for each class, respectively. In [29], Narin et al. have thoroughly investigated the impact of transfer learning on the analysis of chest X-ray radiographs. Five pre-trained models, namely ResNet50, Resnet101, ResNet152, InceptionV3, and Inception-ResNetV2 were the models examined by them and they have achieved accuracies of 96.1%, 99.5%, 99.7% in three different datasets. In addition, Goel et al. [30] have proposed OptCoNet; an optimized Convolutional Neural Network for detecting COVID-19 from X-ray images. They employed the gray wolf optimization algorithm with the aim of tuning the hyperparameters of the classifier and achieved 97.78% accuracy.

Furthermore, more recently ViT-based models have been put forward for COVID-19 detection. In [31], a novel model with two branches has been proposed. In this work, a ViT architecture is utilized as a backbone, integrated with a Siamese network for processing an augmented version of the input X-ray image. They could obtain 99.13% for their accuracy in a 80:20 distribution of train and test. Further, Mondal et al. [32] proposed a ViT-based model and employed a multi-stage transfer learning technique to address the scarcity of data. They obtained an overall accuracy of 96.00%. Furthermore, Liu et al. [33] have applied a transformer-like classifier model. By employing transfer learning techniques in their approach, they improved the results to outperform CNN-based models, achieving 99.7 % accuracy. Additionally, in [34], the authors applied a ViT-based algorithm based on the Swin transformer for feature learning and aggregation in two stages of segmentation and classification. In their paper, they further validated the superiority of their algorithm by comparing their results with well-known visual feature extractors, i.e, EfficientNetV2. The accuracy of 94.3 % was obtained by their approach.

Furthermore, we have provided Table 1, which details an overview of the existing research works on the diagnosis of COVID-19 from CT or X-ray images.

**Table 1.** An overview of the existing works.

| Author, Year | Dataset (CT/X-Ray) | Classification | Approach | Train/Test/validation |
|---|---|---|---|---|
| Konar et al, [35] | X-Ray | Binary | Proposed Semi-Supervised Classifier | Random sampling 70% for training and 30% for testing |
| Vaid et al., [36] | X-Ray | Binary | VGG-19 | Random sampling 80:20:20 for train, validation and testing. |

| Ozturk et al., [37] | X-Ray | Binary and Multi-Class | DarkNet | 5-fold cross-validation |
|---|---|---|---|---|
| Panwar et al., [38] | X-Ray | Binary | Proposed nCOVnet | Random sampling 70% for training and 30% for testing |
| Ahuja et al., [39] | X-Ray | Binary | ResNet-18 | Random sampling. 70% for training and 30% for testing. |
| Sharifrazi et al., [40] | X-Ray | Binary | Sobel + SVM / Proposed Classifier | 10-fold cross-validation |
| Khozeimeh et al., [41] | Numeral | | CNN-AE | 10-fold cross-validation |
| Al Rahhal et al., [31] | CT/X-Ray | Multi-Class | Proposed Siamese+ViT Classifier | 60:40 80:20 :20:80 |
| Mondal et al., [32] | CT/X-Ray | Multi-Class | Proposed xViTCOS + Multistage Transfer Learning | 80:20 |
| Krishnan and Krishnan, [2] | X-Ray | Binary | Pretrained ViT | 73:3:24 |

## 3. Methodology

This section includes our methodology for detecting COVID-19 from X-ray images. The workflow of the adopted pipeline is shown in Figure 3.

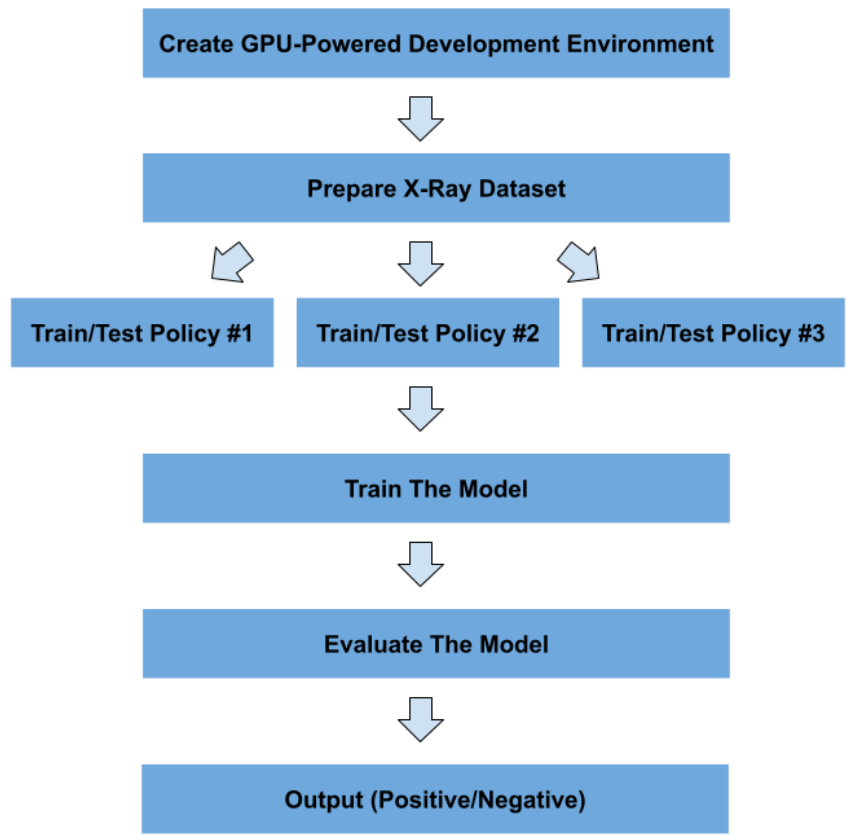

**Figure 3.** Workflow of the proposed pipeline for detecting COVID-19.

Moreover, in this section, after describing the details of the used dataset, all the main components of the proposed method will be elaborated.

### 3.1. Dataset Description

In this paper, a publicly available dataset [1] is used for training and evaluating our proposed method. Table 2 shows the official distribution of this dataset.

**Table 2.** The dataset distribution.

|  | No. Train Samples | No. Test Samples |
|---|---|---|
| Positive (COVID-19) | 16490 | 200 |
| Negative (NO COVID-19) | 13992 | 200 |
| Total | 30482 | 400 |

Moreover, Figure 4 demonstrates some samples from both positive and negative classes.

---

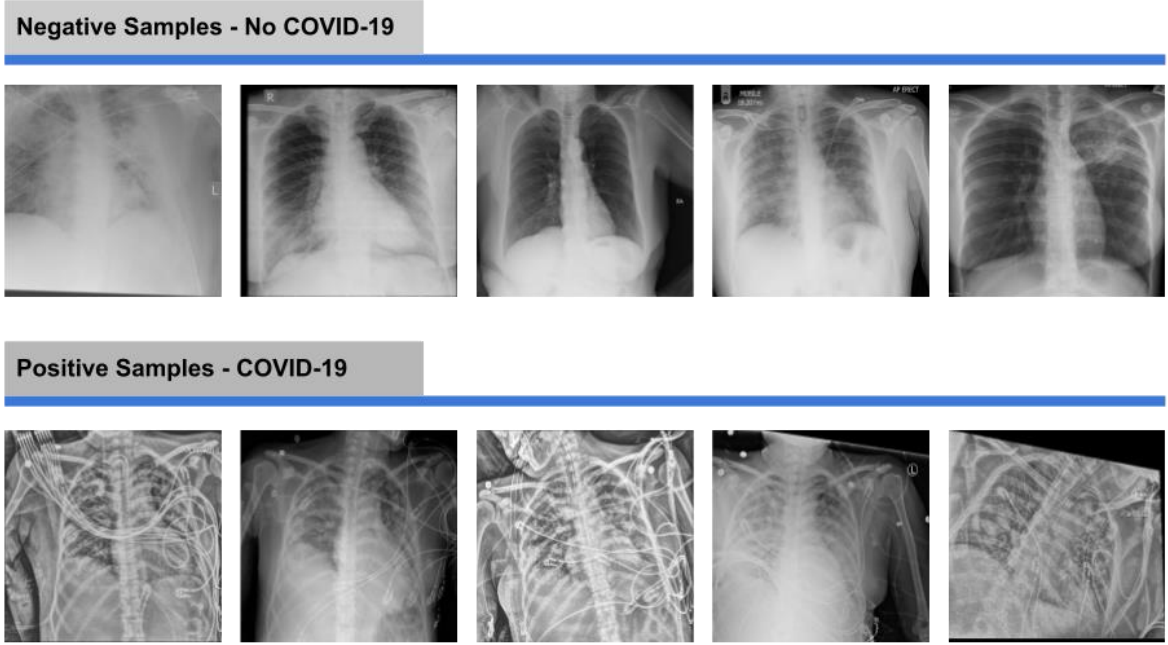

**Figure 4.** Negative and positive samples of X-ray images from the dataset.

### 3.2. Proposed Architecture

Compact Convolutional Transformers or in brief, CCT, develop over Compact Vision Transformers (CVT) and take advantage of a convolutional tokenizer leading to the preservation of local information and production of richer tokens. Compared to the primitive ViT, the convolutional tokenizer is more effective in encoding the connection between patches. In the sequel, we go into further detail on the components of the compact transformers.

As for CCT model design, the original Vision Transformer [23], and original Transformer [42] are proposed. The encoder is made up of transformer blocks, each of which has an MLP block and an MHSA layer. The encoder additionally employs dropout, GELU activation, and Layer Normalization. It is considered that the vision transformers are more compact and simpler. Variants with (the minimum number of) 2 layers, 2 heads, and 128-dimensional hidden layers are implemented. Based on the image resolution of the training dataset, the tokenizers are modified. These variations are referred to as ViT-Lite, and although they differ in size, they are largely comparable to ViT in terms of architecture.

ViT and almost all general transformer-based classifiers follow BERT [43], which sends a learnable class or query token across the network before feeding it to the classifier leading to the conversion of the sequential outputs to a single class index. However, in CCT, an attention-based technique that pools over the output token sequence are leveraged, and hence, unlike the learnable token, the output sequence contains substantial information that includes several parts of the input image, resulting in a more efficient performance. Moreover, the network can correlate data across the input data and weigh the sequential embedding of the transformer encoder's latent space. Finally, Compact Vision Transformer (CVT) is made by substituting SeqPool for the ordinary class token in ViT-Lite.

As for the last steps in designing CCT, a straightforward convolutional block is substituted for the patch and embedding in ViT-Lite and CVT to induce an inductive bias into the model. A single convolution, ReLU activation, and a max pool make up the standard and customary design of this block by which the models have more flexibility than models like ViT since they are no longer restricted to input resolutions that are strictly divisible by the predetermined patch size. The CCT is produced via this convolutional tokenizer, SeqPool, and the transformer encoder. An overview of the proposed architecture has been illustrated in Figure 5.

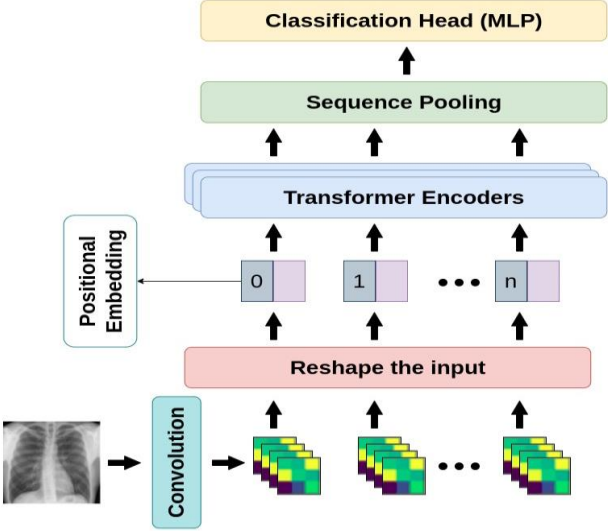

**Figure 5.** An overview of the proposed architecture.

### 3.3. Evaluation Metrics

The measures used for evaluating the performance of the proposed classifier are estimated against the following metrics:

**Confusion Matrix (CM):** A matrix, containing four main elements, namely True Positive (TP), True Negative (TN), False Positive (FP), and False Negative (FN). For a binary classifier, CM is as Figure 6.

**Figure 6.** Confusion Matrix.

**True Positive (TP):** the number of infected samples correctly classified as infected

**True Negative (TN):** the number of uninfected samples correctly classified as uninfected

**False Positive (FP):** the number of infected samples correctly classified as infected

**False Negative (FN):** the number of infected samples correctly classified as infected

Based on the metrics mentioned above, the metrics detailed in Table 3 can be deduced and used for evaluating a classifier.

**Table 3.** The metrics used for evaluation.

| Metric Name | Equation |
|---|---|
| Accuracy | $\dfrac{TP + TN}{FP + FN + TP + TN}$ |
| Precision | $\dfrac{TP}{TP + FP}$ |
| Recall or Sensitivity | $\dfrac{TP}{TP + FN}$ |
| F1-Score | (2*precision * recall) / (precision + recall) |
| AUC-ROC | Area Under Curve of Receiver Operator Characteristic |
| True Positive Rate (TPR) | TP / (TP + FN) |
| False Positive Rate (FPR) | FP / (FP + TN) |
| False Negative Rate (FNR) | FN / (TP + FN) |
| True Negative Rate (TNR) | TN / (TN + FP) |

## 4. Results

This section includes the results of classification by our proposed approach.

### 4.1. Experimental Setup

Table 4 details the software and hardware used for implementing our proposed method.

**Table 4.** Experimental setup.

| | |
|---|---|
| Programming language | Python 3.7 |
| Deep learning library | Pytorch 1.9 |
| CPU | Intel® Core™ i7-10700 CPU @ 2.90GHz × 16 |
| GPU | GeForce GTX 1060 |

### 4.2. Hyperparameter settings

Table 5 details the Hyperparameter settings applied for implementing our proposed method.

**Table 5.** Hyperparameter settings.

| Parameter name | Detail |
|---|---|
| image size | (256, 256) |
| embedding dimension | 512 |
| number of convolution layers | 4 |
| Pooling kernel size | 5 |
| Pooling padding | 1 |
| Pooling stride | 2 |
| kernel size | 5 |
| stride | 2 |
| padding | 1 |
| number of heads | 8 |
| number of classes | 2 |
| positional embedding | Sine Function |

**4.3. Dataset split**

Note that we opted for three main policies for evaluating the classifier. These three are:

1) Policy #1: We used the official training data for training and validating the model and the official test data for testing it.

2) Policy #2: We amalgamated official train and test data with each other; then randomly shuffled the data multiple times. Next, we used 10 fold cross-validation method for the training and evaluation process

3) Policy #3: We randomly shuffled the training data multiple times and then chose a specific number of training data (randomly chosen), removed them from the train set, and added them to the testing set. The number of replaced samples was set in a way to make the test size 0.1 of the remaining training data.

The main reason for pursuing these policies is the small size of the official test chunk, which makes the evaluation results unreliable. This process is depicted in Figure 7.

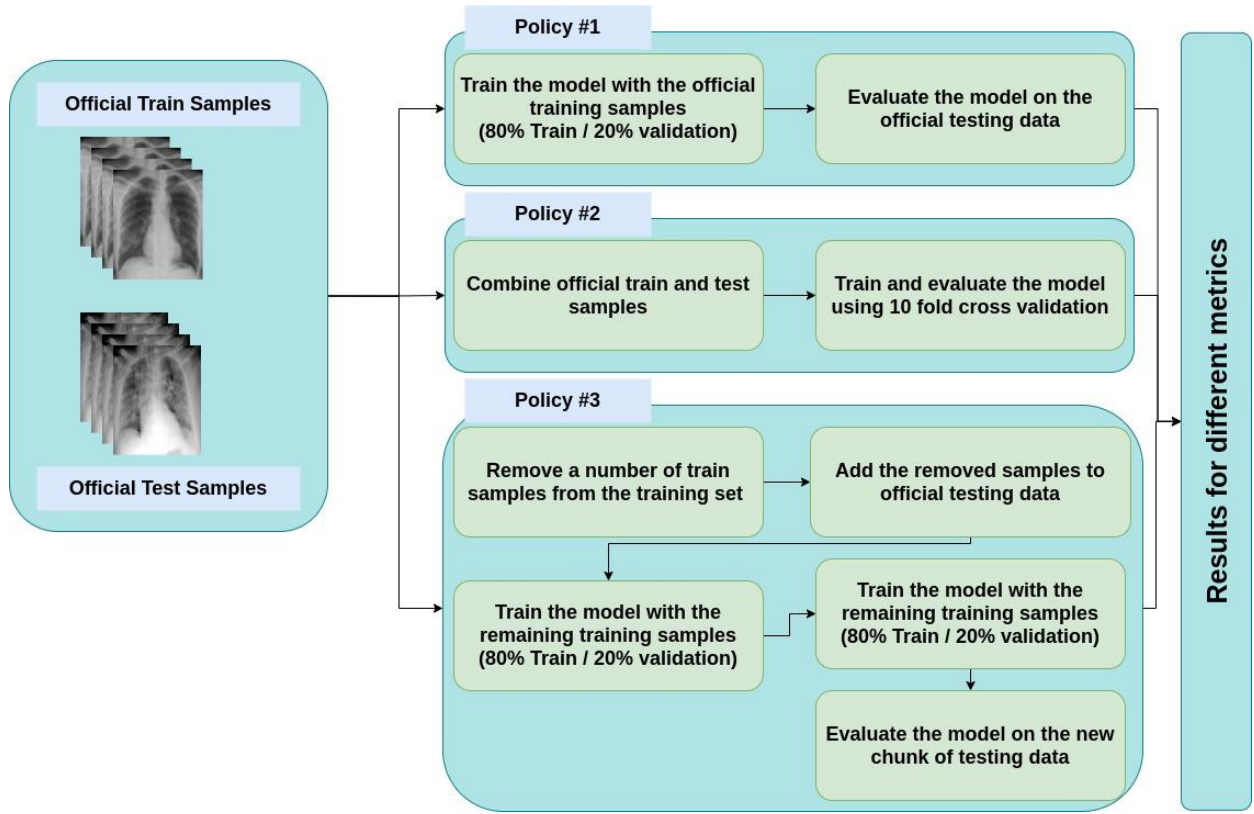

**Figure 7.** Policies for train and test split.

### 4.3.1. Results for Policy #1

Table 6 contains the results of classification by the proposed model on the official distribution of the used dataset. Additionally, Figure 8 shows the obtained CM for the same distribution. Figure 9 and Figure 10 show the accuracy and loss curves vs. epochs, respectively. Figure 11 demonstrates the ROC curve for the classifier.

**Table 6.** Results of classification on the official test data (All metrics are reported on a 0-100 scale).

| Accuracy | Precision | Recall | F1-Score | AUC-ROC | FPR | FNR | TNR |
|----------|-----------|--------|----------|---------|------|------|-------|
| 99.00 | 99.00 | 99.00 | 99.00 | 99.67 | 1.00 | 1.00 | 99.00 |

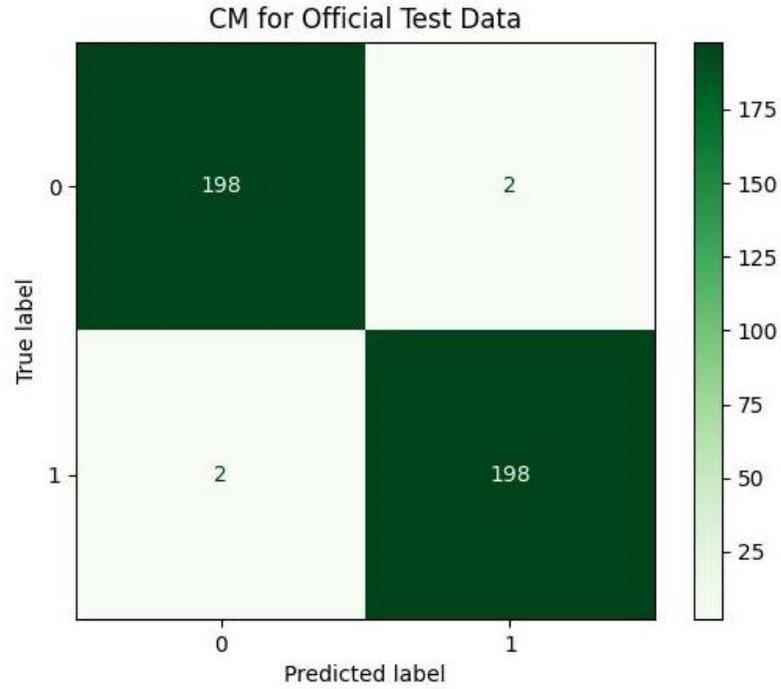

**Figure 8.** Confusion Matrix (CM) for official test data.

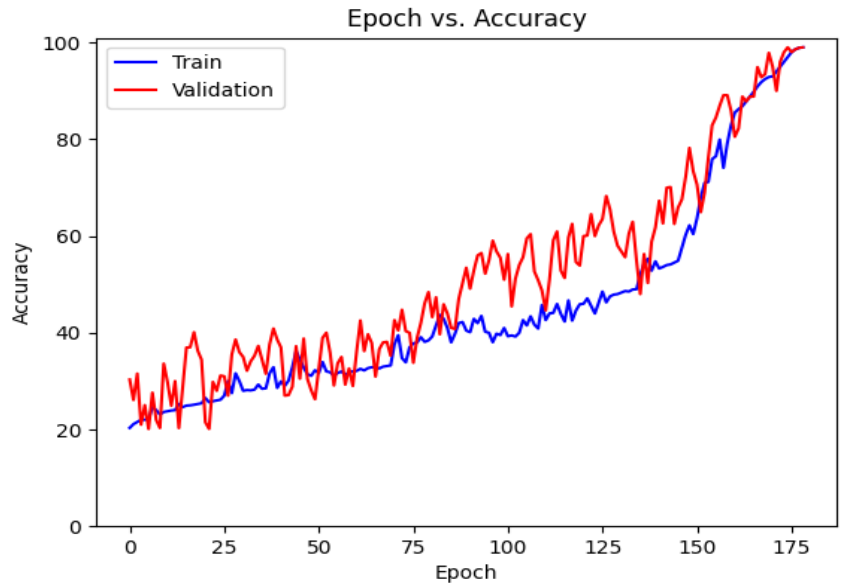

**Figure 9.** Accuracy diagram vs. epochs.

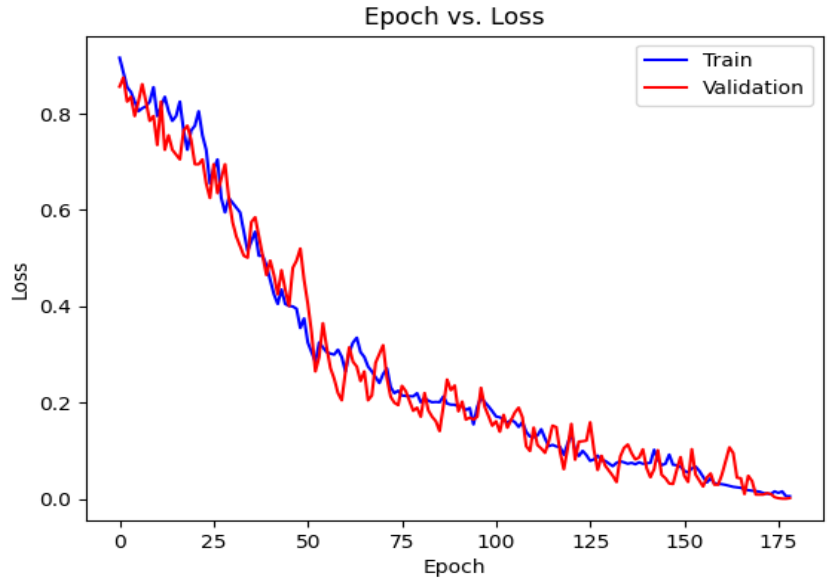

**Figure 10.** Loss diagram vs. epochs.

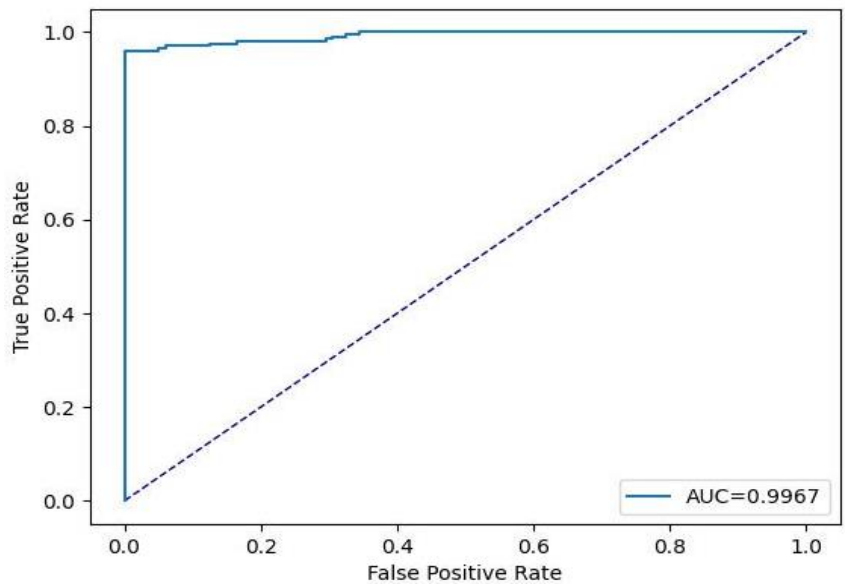

**Figure 11.** Receiver Operating Characteristic (ROC) curve for official test data.

Based on Table 6, it can be observed that our approach can achieve a high value of 99.00 % for accuracy, precision, recall, and F1-Score. The stability of the proposed model in terms of detecting both negative and positive samples can be proved by the fact that a high value of 99.67 is achieved for AUC-ROC.

### 4.3.2. Results for Policy #2

Table 7 details the achieved results for each fold based on the introduced metrics. Figure 12 and Figure 13 demonstrates the accuracy and validations curves achieved in the training process. Figure 14 (a-j) shows the CMs obtained in the second policy.

**Table 7.** Results of classification using 10-fold cross validation (All metrics are reported in 0-100 scale).

| No. Fold | Accuracy | Precision | Recall | F1-Score | AUC-ROC | FPR | FNR | TNR |
|----------|----------|-----------|--------|----------|---------|------|------|-------|
| 1 | 99.16 | 98.90 | 99.42 | 99.16 | 99.23 | 1.10 | 0.58 | 98.90 |
| 2 | 98.90 | 98.21 | 99.61 | 98.91 | 99.15 | 1.81 | 0.39 | 98.19 |
| 3 | 99.71 | 99.68 | 99.74 | 99.71 | 99.20 | 0.32 | 0.26 | 99.68 |
| 4 | 99.38 | 99.10 | 99.68 | 99.39 | 99.61 | 0.91 | 0.32 | 99.09 |
| 5 | 98.96 | 98.40 | 99.55 | 98.97 | 99.03 | 1.62 | 0.45 | 98.38 |
| 6 | 99.61 | 99.42 | 99.81 | 99.61 | 99.27 | 0.58 | 0.19 | 99.42 |
| 7 | 99.03 | 98.84 | 99.22 | 99.03 | 99.19 | 1.17 | 0.78 | 98.83 |
| 8 | 98.80 | 97.90 | 99.74 | 98.81 | 99.02 | 2.14 | 0.26 | 97.86 |
| 9 | 99.64 | 99.68 | 99.61 | 99.64 | 99.64 | 0.32 | 0.39 | 99.68 |
| 10 | 99.03 | 98.65 | 99.42 | 99.03 | 99.31 | 1.36 | 0.58 | 98.64 |
| Average | **99.22** | **98.88** | **99.58** | **99.23** | **99.27** | **1.13** | **0.42** | **98.87** |

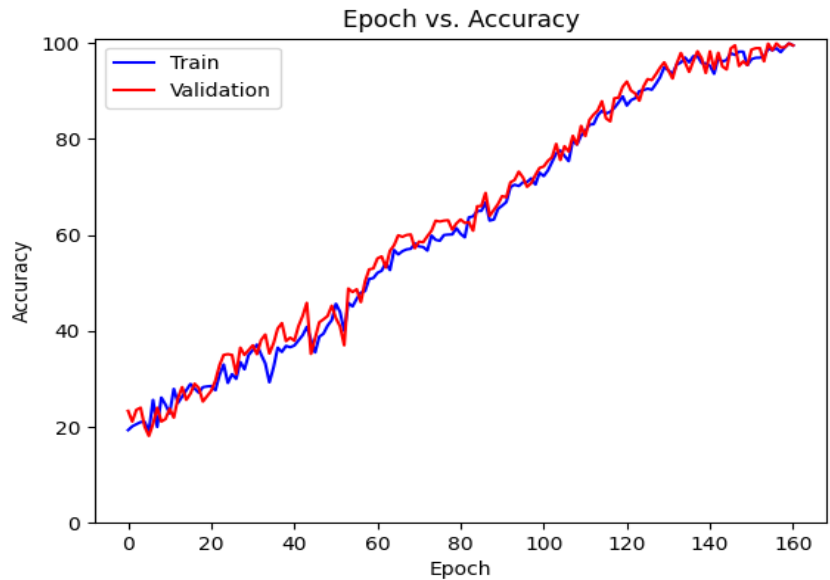

**Figure 12.** Accuracy diagram vs. epochs.

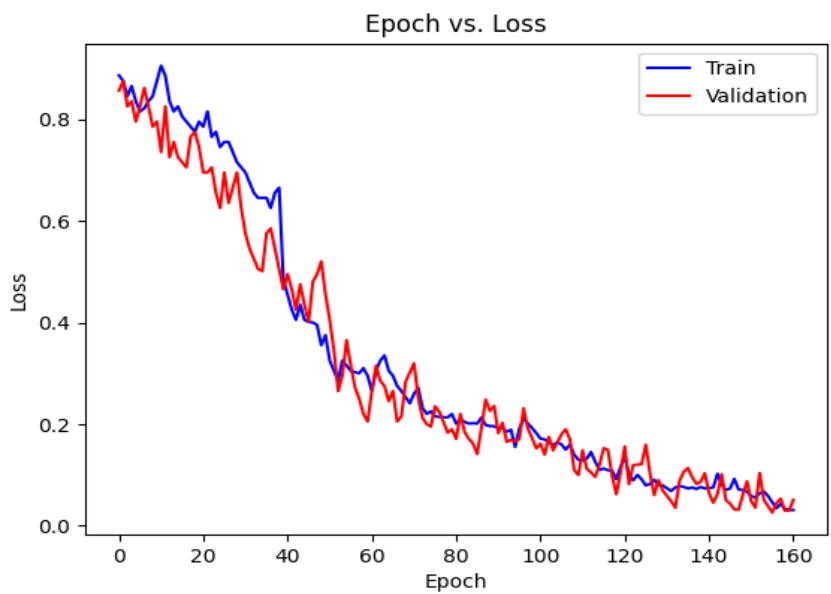

**Figure 13.** Loss diagram vs. epochs.

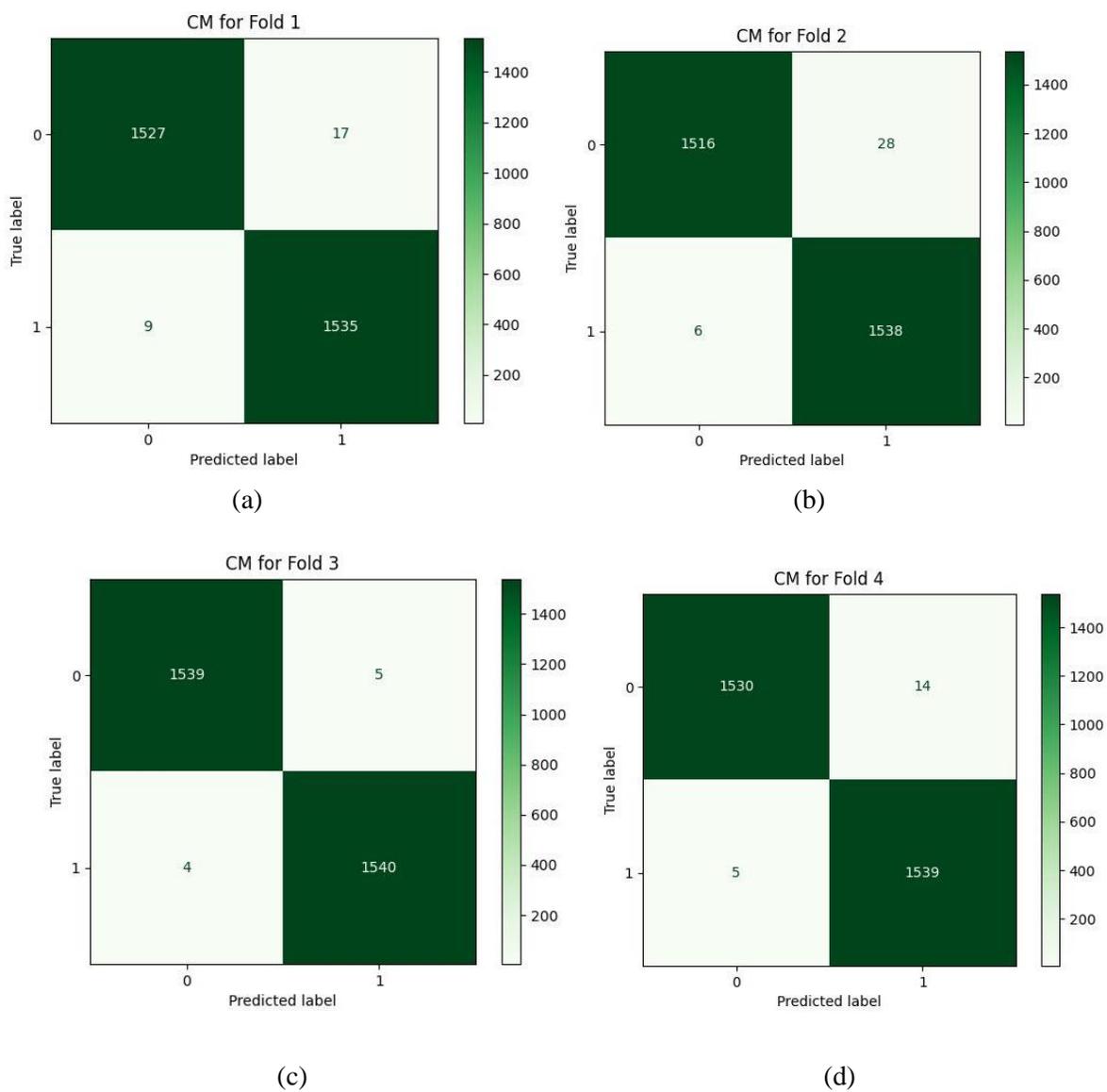

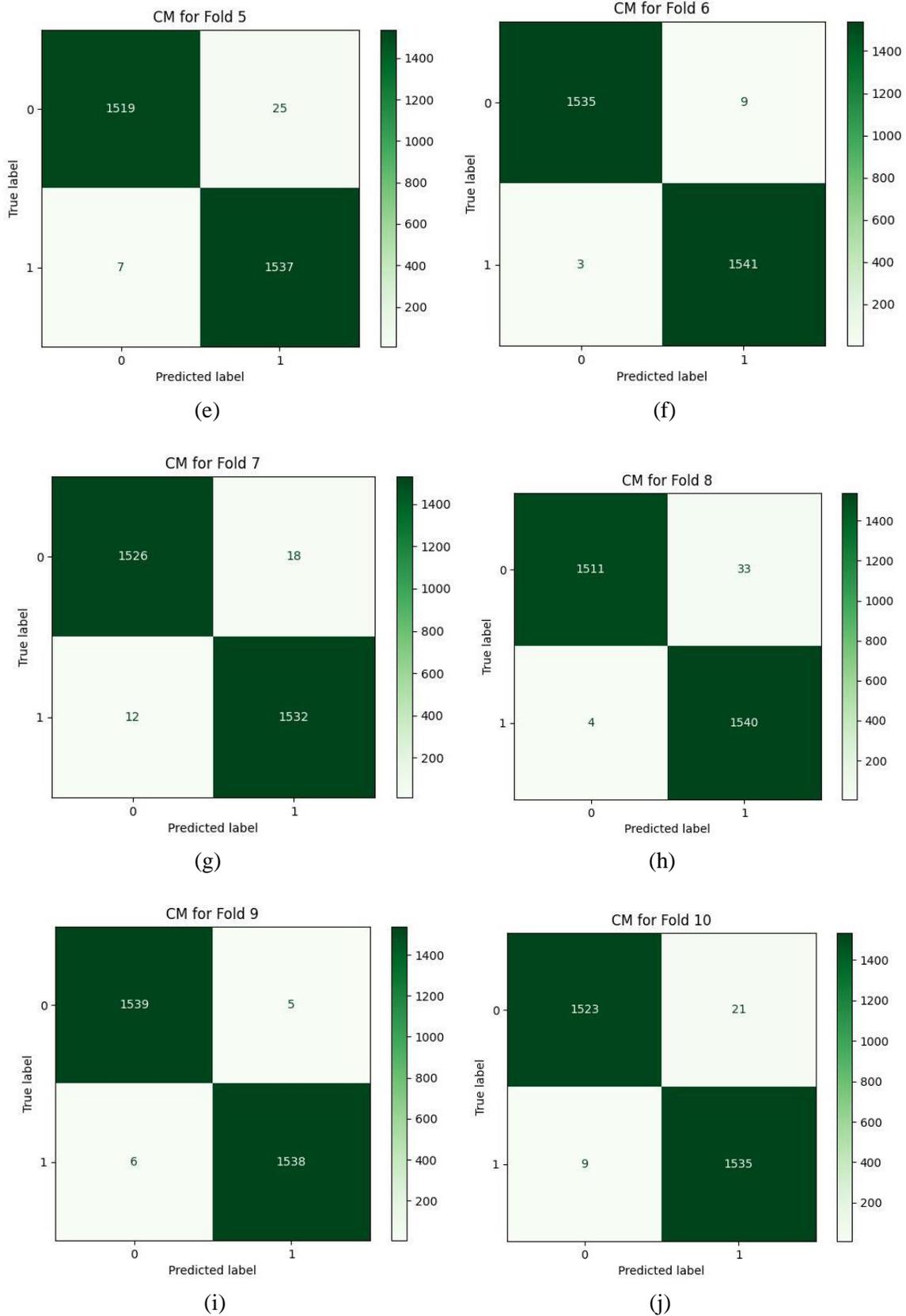

**Figure 14.** Receiver Operating Characteristic (ROC) curve for official test data.

Table 7 shows the results achieved in all folds as well as the average. The achieved accuracy, on average, is 99.22, the precision is 98.88, the recall is 99.58, and the F1-Score is 99.23. The value for AUC-ROC, on average, is 99.27 which shows the strong confidence of the proposed classifier in classifying both negative and positive samples.

### 4.3.3. Results for Policy #3

This subsection includes our results based on the third evaluation policy. Figure 15 demonstrates training and validation accuracy in each epoch. Also, Figure 16 illustrates training and validation loss in the training procedure. Table 8 shows the results achieved by the classifier when we adopt policy 3 for the evaluation. Also, the obtained CM and ROC, in this policy, is shown in Figure 17 and Figure 18, respectively.

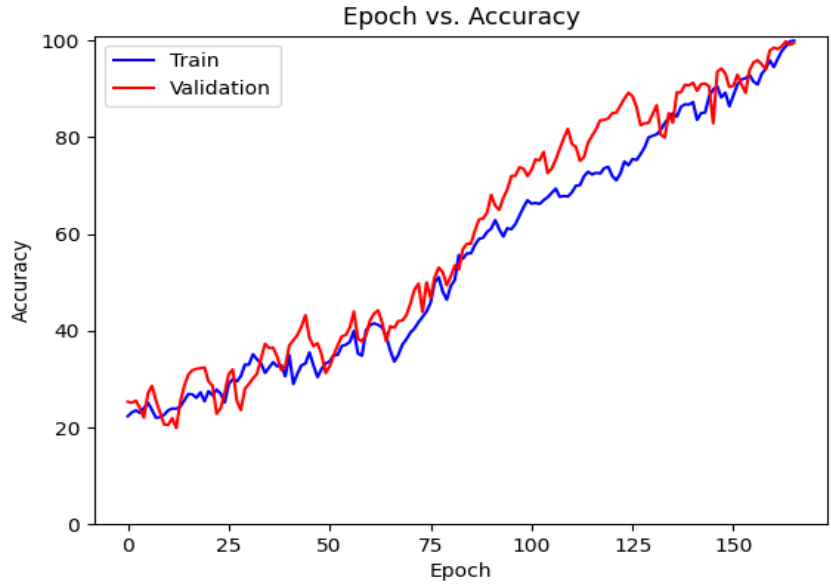

**Figure 15.** Accuracy diagram vs. epochs.

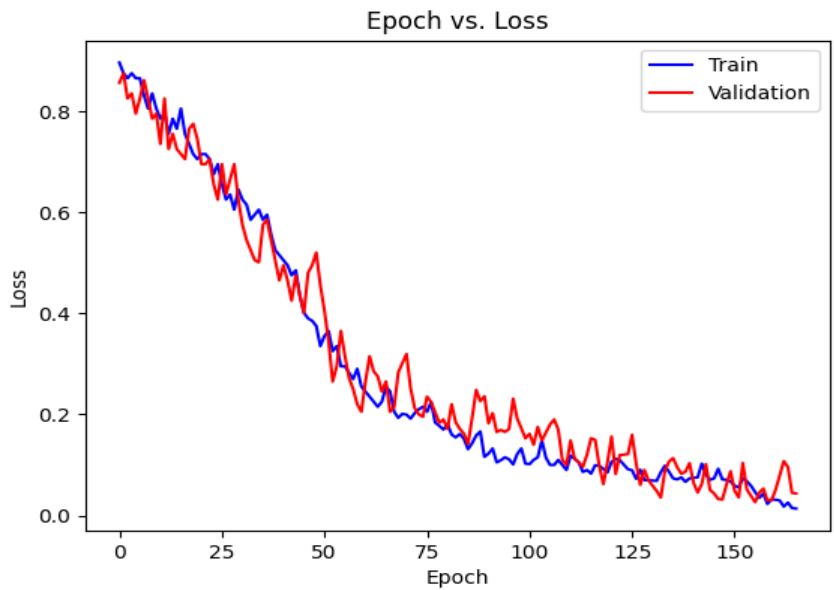

**Figure 16.** Loss diagram vs. epochs.

**Table 8.** Train and test distribution in policy #3.

|  | No. Train Samples | No. Test Samples |
|---|---|---|
| Positive (COVID-19) | 26142 | 4740 |

**Table 9.** Results of classification using policy three (All metrics are reported on a 0-100 scale).

| Accuracy | Precision | Recall | F1-Score | AUC-ROC | FPR | FNR | TNR |
|---|---|---|---|---|---|---|---|
| 99.09 | 98.74 | 99.45 | 99.09 | 99.73 | 1.27 | 0.55 | 98.73 |

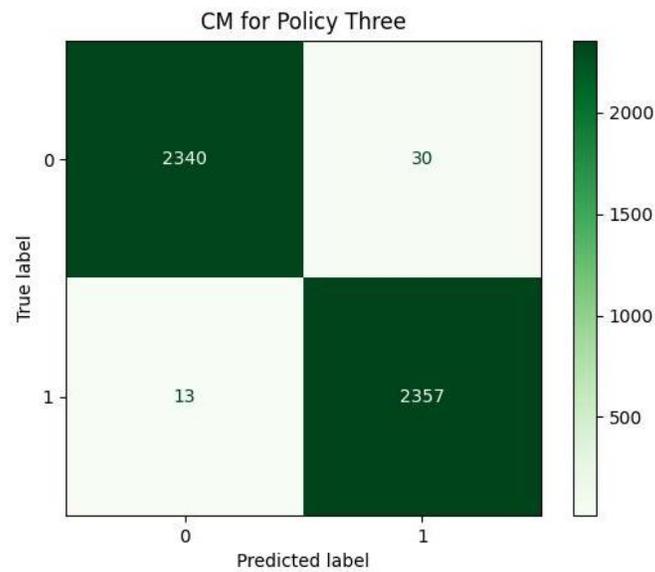

**Figure 17.** Confusion matrix.

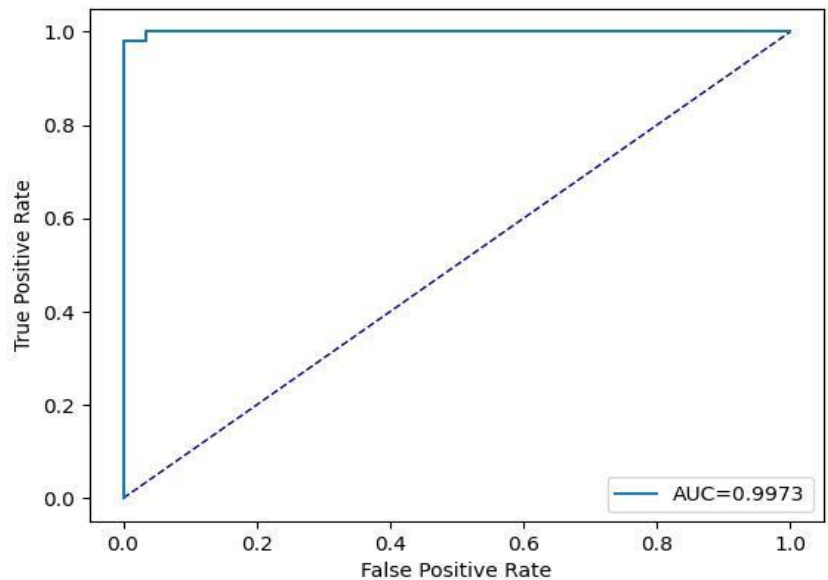

**Figure 18.** Receiver Operating Characteristic curve.

Based on Table 9, we can see that in policy 3, we have achieved 99.09 as accuracy, 98.74 as precision, 99.45 as recall, and 99.09 as F1-Score. 99.73 is achieved as the AUC-ROC of the classifier in policy 3 and proves the efficient performance of the model in distinguishing positive and negative samples correctly.

## 5. Discussion

The COVID-19 serious illness that began in the final months of 2019 and quickly spread to other regions of the world, led to one of the most destructive pandemics. The WHO estimates that as of August 2022, there have been more than 6.4 million deaths and 570 million confirmed cases. According to the research and experiences obtained up to now, CT scans and X-ray images are highly effective tools in diagnosing COVID-19. The absence of specialized human resources in many areas makes it difficult to benefit from such imaging technologies that are encouraged. The scientific community uses computer-aided intelligent systems to automate the desired procedure.

In this study, we proposed deep learning methods for the detection of COVID-19, based on X-ray images of both confirmed (positive) and negative COVID-19 cases that were gathered in a dataset with approximately 31,000 samples. The main architecture that we proposed was CCT. Because of its compactness, CCT can be implemented in low-resource environments, which is its primary advantage, and therefore, is considered to be among mobile-friendly models. In addition, because CCT is a hybrid model based on CNN and ViT, it combines the benefits of both while avoiding their drawbacks. For instance, CCT experiences substantial performance improvements, resulting in a top-1% accuracy of 98% on CIFAR-10. The CCT model is the only transformer-based model among the top 25 models in terms of performance and efficiency on CIFAR-10, despite having no pre-training and being rather small compared to the majority. Moreover, CCT surpasses the majority of comparable CNN-based models in this field, except for some Neural Architectural Search techniques [44]. Furthermore, CCT can be lightweight, using only 0:28 million parameters, while still achieving accuracy within 1% of the top 90% of similar models on CIFAR-10. CCT obtains 80.67% accuracy on ImageNet with fewer parameters and less computational work, and it outperforms more recent, comparably sized models like DeiT [45]. (For more information, see [16]).

In order to achieve better performance in our study, we chose to evaluate the classifier according to three primary policies. In policy 1, we merely trained and validated the model through the official training data, and we examined the classifier using the official test data. Afterward, to obtain more reliable and robust outcomes, the official test and train data were combined, after which they were repeatedly shuffled at random. The training and evaluation process was then conducted using the 10-fold cross-validation method which altogether constitutes our second policy. Finally, to provide the third (and the last) policy, we shuffled the training data at random several times followed by randomly selecting a group of training data, taking them out of the train set, and adding them to the testing set. It is important to note that the official test chunk's small size, which renders the evaluation results untrustworthy, was our main motivation for considering these three different policies and approaches.

Table 10 lists several state-of-the-art research studies on the topic of COVID-19 diagnosis based on X-ray images or CT scans, and the performance of each study is mentioned due to the evaluation metrics used by the authors.

**Table 10.** Comparison.

| Reference | Performance |
|-----------|-------------|
| [46] | Accuracy: 86.66%<br>Precision: 86.75%<br>Recall / Sensitivity / TPR:99.42%<br>F1-Score: 91.89%<br>AUC-ROC:62.50% |

| [47] | Accuracy: 96.296%<br>F1-Score: 95.868%<br>AUC-ROC:0.9821 |
|---|---|
| [48] | Accuracy: 99.78%<br>Precision:1.0<br>Recall / Sensitivity / TPR: 1.0<br>F1-Score: 1.0<br>Specificity: 0.9921% |
| [49] | Accuracy: 92.95%<br>Precision: 91.5%<br>Recall / Sensitivity / TPR: 85%<br>Specificity: 82% |
| [50] | Accuracy: 98.23%<br>Precision:<br>Recall / Sensitivity / TPR: 92.08%<br>F1-Score:<br>Specificity: 99.08% |
| [51] | Accuracy: 97.77%<br>Precision: 97.14%<br>Recall / Sensitivity / TPR:97.14% |
| [52] | Accuracy: 99.93%<br>Precision: 100%<br>Recall / Sensitivity / TPR: 99.90%<br>F1-Score:99.93%<br>Specificity: 100% |
| [53] | Accuracy: 99.02%<br>Recall / Sensitivity / TPR: 98.04%<br>Specificity: 100% |
| [54] | Precision: 0.9301<br>Recall / Sensitivity / TPR: 0.949801<br>F1-Score: 0.939847 |
| [54] | Accuracy: 93.9%<br>Precision:<br>Recall / Sensitivity / TPR: 96.8% |

Following is a brief description of the methodology and results of the articles listed in the table above. In Alakus and Turkoglu's study [46], six different deep learning model types were developed and the outcomes were compared. With an accuracy of 86.66%, LSTM produced the best results out of the group.

In [47], 1345 CT scans were subjected to deep feature extraction using deep learning models like ResNet-50, ResNet-101, AlexNet, etc. Following that, classification methods were given the deep features, and test images were used for model evaluation. The results showed that ResNet-50 and the SVM together provided the best performance. The F1-score was 95.868%, the accuracy was 96.296%, and the AUC was 98.21%.

Srivastava et al. in [48] proposed CoviXNet, a novel CNN-based model, over a dataset of three classes: COVID-19, normal X-rays, and viral-pneumonia infected chest X-ray images, with an accuracy of 99.47% for binary classification (i.e., normal Chest X-ray image and COVID-19 infected), and 96.61% for 3-class classification.

The literature study [49] suggested a CNN-based plus histogram-oriented gradients (HOG) model on a public dataset of 60,000 X-ray images with 400 positive COVID-19 samples and a 92.95% accuracy was attained.

In [50], features from 1125 X-ray images, including 125 images identified as COVID-19 were extracted using DenseNet169. The XGBoost classifier was then fed the derived features and the average accuracy was 98.23%.

A deep learning ResNet50 network was utilized as a classifier in the study [51] to identify viral/bacterial pneumonia and normal cases among 1832 X-ray chest images. Additionally, the ResNet-101 was employed to determine COVID-19 in patients with positive viral-induced pneumonia and the overall accuracy was 97.77%.

A parallel design (COVID-DeepNet) that combines a deep belief network with a convolutional deep belief network trained from scratch on a large dataset was proposed by Al-Waisy et al. [52]. With a 99.93% detection accuracy rate, the method properly identified COVID-19 in patients.

Ten well-known deep learning-based techniques for distinguishing COVID-19 from non-COVID-19 in CT scan images were proposed by Ardakani et al. [53], and the results showed that the models ResNet-101 and Xception (accuracy: 99.02%) achieved the highest overall performance.

To detect COVID-19 infections from chest X-ray images, Mahajan et al. [55] developed a single shot MultiBox detector (SSD) in conjunction with deep transfer learning models and achieved high precision (i.e., 93.01%).

The authors of [54] used transfer learning to diagnose COVID-19 over 1326 chest X-ray images, and the final method, COVID-CXNet, was developed using the well-known CheXNet model [56].

To evaluate the performance of our proposed models, we took into account almost all of the standard and most important evaluation metrics, including accuracy (99.22%), precision (98.88%), recall (99.58%), F1-score (99.23%), AUC-ROC (99.27%), FPR 1.13, FNR (0.42%), and TNR (98.87%), which is outstanding in this regard. The results of our study show that this research is superior to many similar and state-of-the-art works in general and also when each of the evaluation metrics is considered or is completely comparable with them, and Table 10 confirms this claim.

## 6. Conclusion and future works

In this paper, a transformer-based model is proposed for screening chest X-ray images to detect COVID-19 disease. The proposed model is based on Compact Convolutional Transformers, whose main advantage over the other transformer-based models is its less need for data. This is important since in most medical domains data scarcity is ubiquitous. Using different metrics, we have demonstrated the efficacy of the proposed model for COVID-19 diagnosis. In future work, we tend to evaluate our proposed approach to other diseases related to human beings' lungs. That is to say, instead of classifying in a binary fashion for positive and negative COVID-19, the approach should detect more classes of lung disorders.

**Conflict of Interest**

The authors declare that the research was conducted in the absence of any commercial or financial relationships that could be construed as a potential conflict of interest.

**Author Contributions**

AM and JHJ designed the study. AM performed implementation of the approach. Literature review performed by MM. Methodology has been written by AB and MN. Discussion has been done MN. The final version of the article has been edited by JHJ, AB, MAN. MN supervised the project and RL co-supervised the study . All authors have read and approved the final manuscript.

**Funding**

None.

**Acknowledgments**

Not applicable.